\documentstyle[psfig]{article}

\begin{document}
\title{Monte Carlo simulation of quantum computation\thanks{Submitted 
to the IMACS Seminar on Monte Carlo Methods, Brussels, April 1-3, 1997.}}
\author{N. J. Cerf and S. E. Koonin\\
      W. K. Kellogg Radiation Laboratory, 106-38\\
      California Institute of Technology,
      Pasadena, California 91125}

\date{January 1997}

\maketitle
\begin{abstract}
The many-body dynamics of a quantum computer can be reduced
to the time evolution of non-interacting quantum bits in auxiliary fields 
by use of the Hubbard-Stratonovich representation of two-bit
quantum gates in terms of one-bit gates. This makes it possible 
to perform the stochastic simulation of a quantum algorithm,
based on the Monte Carlo evaluation of an integral of dimension
polynomial in the number of quantum bits.
As an example, the simulation of the quantum circuit 
for the Fast Fourier Transform is discussed.
\end{abstract}

\vskip 1 cm

\section{Introduction}
The potential use of quantum computers for solving certain
classes of problems has recently received a considerable amount
of attention (see, {\it e.g.},
\cite{bib_rev_lloyd,bib_rev_divincenzo,bib_ekertjozsa}
for a comprehensive review).
Several quantum algorithms have been developed, such as quantum 
factoring~\cite{bib_shor}, having the potential for revolutionizing
computer science. The purpose of this paper is to explore the application
of a Monte Carlo method that has been developed
in the context of quantum many-body systems
to the simulation of quantum computers~\cite{bib_sek}.
Quantum computers can be seen as peculiar quantum many-body systems
that evolve according to a non-local time-dependent interaction
so as to carry out a ``computation''.
The component quantum bits (qubits)
interact via a sequence of quantum gates, performing each
a prescribed unitary transformation (rotation, Hadamard transformation,
controlled {\sc not}, controlled phase, etc.)~\cite{bib_barenco}.
Two-bit (or $n$-bit) gates
therefore effect {\it non-local} interactions between qubits,
and the ``quantum algorithm'' (characterized by a network of quantum gates)
corresponds to a specific sequence of unitary
transformations, {\it i.e.}, a {\it time-dependent} interaction.
Numerous methods have been developed for years in order to treat general
quantum many-particle systems (see, {\it e.g.}, \cite{bib_negele}).
It is therefore intriguing to examine whether 
the application of the same methods to quantum computers
might be similarly successful. We focus here on
a stochastic approach based on the 
Hubbard-Stratonovich transformation~\cite{bib_hubb}
which has been shown to be suitable for the description of quantum many-body
systems (see, {\it e.g.}, \cite{bib_alhassid,bib_johnson}).
The central idea of this approach is to replace
the many-body propagator for the entire quantum computer
(of say $L$ ``interacting'' qubits) with $L$ one-bit propagators
in fluctuating auxiliary fields, thereby ``decoupling'' the qubits.
More specifically, solving the quantum dynamics of the $L$-bit computer
in a very high-dimensional Hilbert space ($d=2^L$)
reduces to evaluating a high-dimensional -- but polynomial in $L$ --
integral over auxiliary fields. The latter
is then approximated by use of a stochastic method.
\par

\section{Quantum computer as a many-particle system}
\begin{figure}
\caption{Hubbard-Stratonovich representation of a two-bit quantum gate
in terms of two one-bit gates in fluctuating 
auxiliary fields $\sigma$, $\tau$.}
\vskip 0.25cm
\centerline{\psfig{figure=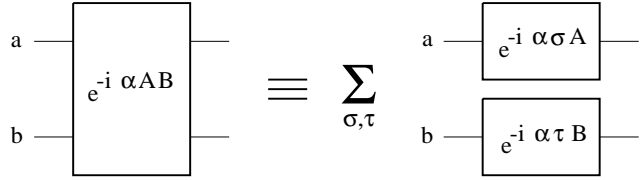,width=3.3in,angle=-90}}
\label{fig_picture}
\vskip -0.25cm
\end{figure}

Consider a quantum computer consisting of a register of
$L$ qubits supplemented with a quantum algorithm,
defined as a sequence of $G$ quantum gates.
The total unitary transformation characterizing the quantum computation
is thus expressed as an ordered product of operators
(note the product from right to left)
\begin{equation}
U \equiv  \prod_{g=1}^G U_g = U_G \cdots U_1
\end{equation}
where $U_g$ is the unitary transformation performed by the $g$-th
gate, and $G$ is the total number of gates. It has been shown that
{\it two-bit} gates are universal, that is quantum gates operating on
one and two qubits are sufficient to construct a general quantum
circuit~\cite{bib_barenco,bib_divincenzo,bib_deutsch,bib_lloyd}.
Therefore, we restrict ourselves to the simulation of quantum circuits made
of $G$ two-bit gates ($U_g$ being a two-bit gate acting
on qubits $a_g$ and $b_g$), keeping in mind that an arbitrary
quantum computation can be achieved with an appropriate sequence of such gates.
(Obviously, one-bit gates can always be incorporated into two-bit gates.)
Note that an efficient quantum algorithm must have $G$ polynomial
in $L$. For example, the quantum Fast Fourier Transform (FFT)
circuit~\cite{bib_coppersmith} used in quantum factoring~\cite{bib_shor}
requires $G=L(L-1)/2$
two-bit gates. A two-bit quantum gate that effects a unitary transformation
on qubits $a_g$ and $b_g$ can be written generically 
as the two-bit operator
\begin{equation}
U_g = e^{-i\alpha_g A_g B_g}
\end{equation}
where $\alpha_g$ is a real number and $A_g$, $B_g$ are two {\it commuting}
one-bit Hermitian operators referring to the qubits involved
in the quantum gate [{\it i.e.}, the operator $A_g$ ($B_g$) 
affects qubit $a_g$ ($b_g$)].
For example, the controlled-{\sc not} gate~\cite{bib_barenco}
acting on qubit $a$ (as a control)
and qubit $b$ (as a target) has $\alpha=\pi/4$, $A=1-\sigma_z$,
and $B=1-\sigma_x$, with $\sigma_x$, $\sigma_z$ being Pauli matrices.
The Hubbard-Stratonovich representation of $U_g$
is obtained by writing the identity
\begin{equation}  \label{eq_square}
i\alpha_g A_g B_g = i\alpha_g (A_g-\tau_g) (B_g- \sigma_g) +
i\alpha_g\sigma_g A_g + i\alpha_g\tau_g B_g - i \alpha_g\sigma_g\tau_g
\end{equation}
where $\sigma_g$ and $\tau_g$ are {\it real} auxiliary fields
corresponding to the $g$-th gate, and then 
integrating the exponential of Eq.~(\ref{eq_square})
over $\sigma_g$ and $\tau_g$, resulting in
\begin{equation}
U_g = \frac{|\alpha_g|}{2\pi} \int_{-\infty}^{\infty} d\sigma_g \; d\tau_g \;
e^{i\alpha_g\sigma_g\tau_g} 
e^{-i\alpha_g\sigma_g A_g} 
e^{-i\alpha_g\tau_g B_g}
\end{equation}
This expression is most important because the {\it two-bit} gate $U_g$
is represented as an infinite sum of products of (field-dependent)
{\it one-bit} gates, 
$e^{-i\alpha_g\sigma_g A_g}$ and $e^{-i\alpha_g\tau_g B_g}$.
For a given value of the fields $\sigma_g$, $\tau_g$, 
the two qubits $a_g$ and $b_g$ act as non-interacting particles
and evolve independently (they do {\it not} become entangled
when initially prepared in a product state).\footnote{Several alternative
Hubbard-Stratonovich representations of a 2-bit quantum gate requiring
only {\em one} auxiliary field per gate can be written, the drawback being
that the involved 1-bit transformations are in general non-unitary.
This is under current investigation.}
Only the sum over fields creates a ``coupling'' between them,
as pictured in Fig.~\ref{fig_picture}.
As a consequence, for a given set of $\sigma_g$'s and $\tau_g$'s,
the time evolution effected by the whole quantum circuit
can be computed separately for each qubit: one calculates the
time-evolution of $L$ qubits in $L$ two-dimensional Hilbert spaces
rather than the time-evolution of a single quantum state (the state of
the entire computer) in the full $2^L$-dimensional Hilbert space. This 
exponential reduction of the size of the Hilbert space appears
clearly when writing the total unitary transformation for the quantum circuit
\begin{equation}  \label{eq_Utot}
U = \int D\sigma \;
\exp \left( {i \sum_g \alpha_g\sigma_g\tau_g} \right) \; 
\underbrace{ \prod_g V_g(\sigma_g)\; W_g(\tau_g) }_{
\displaystyle U[\sigma]}
\end{equation}
where $D\sigma = \prod_g  \frac{|\alpha_g|}{2\pi} d\sigma_g d\tau_g$
is the measure over the auxiliary fields $\sigma_1,\cdots\sigma_G$
and $\tau_1,\cdots\tau_G$, and $U[\sigma]$ is the total unitary
transformation for a given ``path'' $\sigma$ in auxiliary-field space.
Here, $V_g(\sigma_g) \equiv e^{-i\alpha_g\sigma_g A_g}$ and
$W_g(\tau_g) \equiv e^{-i\alpha_g\tau_g B_g}$ stand for
the unitary transformation performed by the one-bit gate
acting separately on qubit $a_g$ and $b_g$, respectively.
(The one-bit gates $V_g$ and $W_g$ replacing the $g$-th
two-bit gate $U_g$ depend respectively on the auxiliary field $\sigma_g$ 
and $\tau_g$.) Eq.~(\ref{eq_Utot}) involves
only {\it one-bit} operators, and therefore describes
the time-evolution of $L$ {\it non-interacting} qubits
(averaged over auxiliary fields). The operator 
$U[\sigma]$ is more conveniently written as a product of one-bit
operators over the $L$ qubits,
\begin{equation}   \label{eq_Usigmatau}
U[\sigma] = \prod_{l=1}^L U^{[l]}[\sigma]
\end{equation}
where the one-bit operator $U^{[l]}$, describing the overall evolution
of the $l$-th qubit, is expressed as the ordered product
\begin{equation}  \label{eq_Usigmatau2}
U^{[l]}[\sigma] = \prod_g U_g^{[l]}(\sigma_g,\tau_g) 
\end{equation}
with
\begin{equation}   \label{eq_Usigmatau3}
U_g^{[l]}(\sigma_g,\tau_g) = \left\{ \begin{array}{l@{\qquad}l}
V_g(\sigma_g) & {\rm if~}l=a_g \\
W_g(\tau_g)   & {\rm if~}l=b_g \\
1 & {\rm otherwise} \;.
\end{array} \right.
\end{equation}
The drawback of this exponential reduction in the Hilbert space
is obviously the $2G$-dimensional integral over fields in Eq.~(\ref{eq_Utot}),
which can only be approximated by a numerical method in general.
The underlying idea of a stochastic method
is to compute only the dominant terms in this integral,
that is to consider the paths in auxiliary-field space 
that contribute the most to it, assuming that this yields
a good estimate of the exact integral.
Several (more or less efficient) Monte Carlo techniques can be thought of
for sampling these paths, but a generic ``sign'' problem unavoidably
occurs due to the complex weight in Eq.~(\ref{eq_Utot}).
This will be discussed later on.
However, the central point here is that the dimension $2G$ of this integral
is polynomial in the dimension of the problem -- {\it i.e.}, polynomial in
the number of qubits $L$ --
at the condition that $G={\rm poly}(L)$. The latter condition is 
fulfilled for any efficient quantum algorithm, suggesting that the Monte Carlo
simulation of a quantum computer might be interesting
if the ``sign'' problem is circumvented.
\par

\section{Stochastic simulation of a quantum computer}
Consider now the stochastic calculation of the quantities of interest
in a general quantum computation. In the context of quantum many-body systems,
stochastic methods are especially appropriate for calculating quantum
expectation values, so that our goal is to
express the output of the quantum computation as an observable.
Assume that the entire quantum computer is initially in a product state
$|0_1 0_2 \cdots 0_L \rangle$. (If this is not the case, the first step
of the computation should simply be the preparation of the correct
initial state from $|0_1 0_2\cdots 0_L \rangle$.) The quantum computation 
({\it i.e.}, the unitary transformation $U$) is implemented by
a quantum circuit acting on this initial state.
The final step of a quantum algorithm is then
to measure a set of ``output'' qubits (not necessarily all the
$L$ qubits). We restrict ourselves here to quantum algorithms that
provide a {\it deterministic} result (unlike quantum factoring).
We assume therefore that the output bits are in a product
state, so they can be measured individually ({\it i.e.}, one can perform
an {\it inclusive} measurement of each of them separately). 
The most general observable $O$ with vanishing variance (deterministic output)
consists then in a product of one-bit observables,
and several such $O$'s can be measured simultaneously.
(If the output bits are not in a product state, one should extend
the quantum computation with a unitary transformation mapping
the entangled final state into a product state.\footnote{It is not
clear whether this requirement makes the extended quantum computation
much harder in a general case. At least, some quantum algorithms are
known to provide a deterministic result, such as Grover's 
algorithm~\cite{bib_grover},
so that the output bits are then in a product state. Note that
the same requirement must be met
for the recently suggested realization of quantum computers
using NMR experiments~\cite{bib_chuang}.})
More generally, the quantum many-particle simulation approach
allows us to prescribe the value of certain qubits in the output register.
We separate the $L$ output qubits
into $L_m$ measured qubits, $L_p$ prescribed qubits, and
$L_t=L-L_m-L_p$ traced over qubits ({\it i.e.}, ``scratch'' qubits
that are necessary to make the overall computation unitary, 
but are not observed in the final measurement).
The observable can then be written as 
\begin{equation}
O=\prod_{\{m\}}^{L_m} O_m
\end{equation}
where $O_m$ is a one-bit observable acting on qubit $m$.
Consequently, the result of a quantum computation can be written
as the expectation value of the observable $O$,
\begin{equation}  \label{eq_expectation}
\langle O \rangle = 
\frac {\langle 0_1\cdots 0_L | U^{\dagger} O P U | 0_1\cdots 0_L \rangle}
      {\langle 0_1\cdots 0_L | U^{\dagger} P U | 0_1\cdots 0_L\rangle}
\end{equation}
where we define the projector $P$ as
\begin{equation}
P=\prod_{\{p\}}^{L_p} P_p
\end{equation}
where $P_p \equiv |\pi_p\rangle \langle \pi_p|$ is a projector on
the prescribed value $\pi_p$ for qubit $p$.
Note that it is crucial to consider a quantum algorithm such that the variance
of $O$ vanishes when the prescribed qubits have the correct value,
so that Eq.~(\ref{eq_expectation}) yields the {\it deterministic} output
of the quantum computation. The only variance in the simulated output
will be the statistical noise resulting from the stochastic evaluation
of Eq.~(\ref{eq_Utot}).
\par

The central point now is that, using Eqs.~(\ref{eq_Utot})
and (\ref{eq_Usigmatau}),
the numerator and denominator in Eq.~(\ref{eq_expectation}) can be 
expressed in terms of an infinite sum of products of $L$ 
one-bit matrix elements for each qubit, 
\begin{equation}
\langle O \rangle = \frac 
{ \int D\sigma \; D\sigma' \; 
e^{i \sum_g \alpha_g (\sigma_g \tau_g - \sigma_g' \tau_g')}
\prod_{l=1}^L \langle 0_l | U^{\dagger [l]}[\sigma'] \, O^{[l]} \, P^{[l]} \,
                      U^{[l]}[\sigma] \, | 0_l \rangle }
{ \int D\sigma \; D\sigma' \; 
e^{i \sum_g \alpha_g (\sigma_g \tau_g - \sigma_g' \tau_g')}
\prod_{l=1}^L \langle 0_l | U^{\dagger [l]}[\sigma'] \, P^{[l]} \,
                      U^{[l]}[\sigma] \, | 0_l \rangle }
\end{equation}
where $\sigma'$ represents the set of auxiliary fields
($\sigma_g'$ and $\tau_g'$)
used in the Hubbard-Stratonovich expression of $U^{\dagger}$.
This can be written more concisely as
\begin{equation}   \label{eq_expecO}
\langle O \rangle = \frac 
{ \int D\sigma \; D\sigma' \; 
         \exp \left( {-iS[\sigma,\sigma']}\right) \; O[\sigma,\sigma'] }
{ \int D\sigma \; D\sigma' \;
         \exp \left( {-iS[\sigma,\sigma']}\right)  }
\end{equation}
where the (complex) action is defined as
\begin{eqnarray}  \label{eq_action}
\lefteqn{ S[\sigma,\sigma'] = 
- \sum_g \alpha_g (\sigma_g \tau_g - \sigma_g' \tau_g' )  } 
\hspace {1.5truecm} \nonumber \\
&+& i \sum_{l=1}^L
\ln \langle 0_l | U^{\dagger [l]}[\sigma'] \, P^{[l]} \,
U^{[l]}[\sigma] \, | 0_l \rangle 
\end{eqnarray}
The operator $P^{[l]}$ is a
one-bit projector if the $l$-th qubit is prescribed,
and the unit operator otherwise. The estimator of $O$ is
\begin{equation}  \label{eq_Osigma}
O[\sigma,\sigma']=
\prod_{l=1}^L 
\frac { \langle 0_l | U^{\dagger [l]}[\sigma'] \, O^{[l]} \, P^{[l]} \,
U^{[l]}[\sigma] \, | 0_l \rangle }
{ \langle 0_l | U^{\dagger [l]}[\sigma'] \, P^{[l]} \,
U^{[l]}[\sigma] \, | 0_l \rangle }
\end{equation}
where $O^{[l]}$ is the one-bit component of the observable $O$ if
the $l$-th qubit is measured, and the unit operator otherwise.
Note that the $L$ matrix elements in the right-hand side
of Eq.~(\ref{eq_action}) are for single qubits, so that
the calculation of the action involves $\sim 4G$ products of
non-unit 2x2-matrices. (There are 2 fields per gate, and the
Hermitian conjugate $U^{\dagger}$ must be considered together with $U$.)
The calculation of Eq.~(\ref{eq_Osigma}) requires essentially the same
operations.
\par

\section{Sampling of the auxiliary-field paths}
Let us now consider the stochastic evaluation of Eq.~(\ref{eq_expecO})
based on a sampling of the paths (set of $\sigma_g$'s ,$\tau_g$'s)
that contribute the most to the integral.
The simplest possibility is to perform an importance sampling
of the paths according to the weight $|e^{-iS}|$. (Note that this
weight is not equal to 1 since $S$ is generally complex.)
This can be done for example using the 
Metropolis method~\cite{bib_metropolis}. A random walk 
in the auxiliary-field space is simulated, such that the limit distribution
of sampled paths is proportional to $|e^{-iS}|$. This makes it possible
to write $\langle O \rangle$ as a ratio of Monte Carlo averages,
\begin{equation}  \label{eq_metrop}
\langle O \rangle \sim \frac  
{ \langle e^{-i\, {\rm Re}\, S[\sigma,\sigma']} \, O[\sigma,\sigma']
   \rangle_{\sigma,\sigma'} }
{ \langle e^{-i\, {\rm Re}\, S[\sigma,\sigma']} \rangle_{\sigma,\sigma'} }
\end{equation}
where $\langle\cdot \rangle_{\sigma,\sigma'}$ stands for the 
simulation average over auxiliary-field paths.
A test of this approach has been carried out,
showing that the term $e^{-i\, {\rm Re}\, S}$ generally makes
the (averaged) numerator and
denominator of Eq.~(\ref{eq_metrop}) exceedingly small.
Unless this ``sign'' problem can be overcome, the standard Metropolis method
seems therefore to be inefficient in this context.\footnote{This 
numerical test has been performed on 
a small quantum circuit ($L=3$, $G=4$) using a one-field
per gate Hubbard-Stratonovich transformation, but we are confident
that the ``sign'' problem is generic.}
Since the weight of the paths in Eq.~(\ref{eq_expecO}) is complex
(this is at the heart of the sign problem)
a more promising possibility is the recourse to a simulation
based on the complex Langevin equation~\cite{bib_parisi,bib_klauder}.
In the Langevin algorithm (see, {\it e.g.}, \cite{bib_okano,bib_adami}),
paths distributed according to the ``complex probability distribution''
$\sim e^{-iS}$ can be generated, allowing
the computation of Eq.~(\ref{eq_expecO}) as a time average
over a guided random walk for the fields in {\it complex} plane.
In the case of interest
here, the random walk for field $\sigma_g$ is the solution
of the stochastic differential equation
\begin{equation} \label{eq_langevin}
\frac {d\sigma_g}{dt} = - \frac{i}{2} \frac{\partial S}{\partial \sigma_g}
+ \eta_g(t)
\end{equation}
where $t$ is a fictitious time (simulation time) 
and $\eta_g$ is a (real) Gaussian white noise
satisfying $\langle \eta_g(t) \rangle =0$ and 
$\langle \eta_g(t)\eta_g(t') \rangle =\delta(t-t')$. The first
term in the right-hand side of Eq.~(\ref{eq_langevin}) can be seen as
a ``string'' force which keeps $\sigma_g$ close to the value
for which the action $iS$ is minimum, while the ``noise'' term is responsible
for the sampling of a region in auxiliary-field space around this extremum.
Although a general proof of the convergence of the complex 
Langevin simulation does not exist~\cite{bib_okano},
it turns out to work very nicely
for a number of systems (the convergence is related to the location
of the repulsive points of the Langevin dynamics).
The Langevin simulation yields then a stochastic estimate
of the output of the quantum computer,
\begin{equation}
\langle O \rangle \sim {1 \over T} \int_{t}^{t+T} dt \;
 O[\sigma(t),\sigma'(t)]
\end{equation}
which is calculated by averaging $O[\sigma,\sigma']$
for a sufficiently long random walk.
Using Eq.~(\ref{eq_action}), the time derivative of the field $\sigma_g$
can be written explicitly as
\begin{equation}  \label{eq_explicitderiv}
\frac {d\sigma_g}{dt} = \frac{i}{2} \alpha_g \tau_g 
+ \frac{1}{2} R_g[\sigma,\sigma'] +\eta_g(t)
\end{equation}
with
\begin{equation}  \label{eq_R}
R_g[\sigma,\sigma']=  
\prod_{l=1}^L \frac { \langle 0_l | U^{\dagger [l]}[\sigma'] \, P^{[l]} \,
d U^{[l]}[\sigma]/d \sigma_g \,  | 0_l \rangle }
{ \langle 0_l | U^{\dagger [l]}[\sigma'] \, P^{[l]} \,
U^{[l]}[\sigma] \, | 0_l \rangle }
\end{equation}
One single term $(l=a_g)$ differs from one in this product as
only the 1-bit gate acting on qubit $a_g$
depends on $\sigma_g$ [see Eqs.~(\ref{eq_Usigmatau2}) and 
(\ref{eq_Usigmatau3})]. One has
\begin{equation}  \label{eq_dUdsigma1}
\frac{dU^{[a_g]}[\sigma]}{d\sigma_g} 
= \prod_{g'} {\tilde U}_{g'}^{[a_g]}(\sigma_{g'},\tau_{g'}) 
\end{equation}
with
\begin{equation}  \label{eq_dUdsigma2}
{\tilde U}_{g'}^{[a_g]}(\sigma_{g'},\tau_{g'}) 
= \left\{ \begin{array}{l@{\qquad}l}
-i\alpha_g A_g U_{g}^{[a_g]}(\sigma_g,\tau_g) & {\rm if~}g'=g \\
U_{g'}^{[a_g]}(\sigma_{g'},\tau_{g'}) & {\rm otherwise}
\end{array} \right.
\end{equation}
The calculation of the derivative $d\sigma_g / dt$
(necessary to increment the fields along the random walk) requires thus
the estimate of $R_g[\sigma,\sigma']$ which is
of the same kind as expression~(\ref{eq_Osigma}) for the observable $O$:
rather than inserting the observable $O$, one inserts the operator $A_g$
(conjugate to the field $\sigma_g$) at a specific point in the ordered
product of propagators. The coupling in the time evolution of the fields
is obvious from Eq.~(\ref{eq_explicitderiv}). In particular,
each pair of fields $(\sigma_g,\tau_g)$ is strongly coupled
through the first term in the right-hand side of 
Eq.~(\ref{eq_explicitderiv}). Indeed, combining Eq.~(\ref{eq_explicitderiv})
and its counterpart for $\tau_g$, it is easy to see that the
time-evolution of the field $\sigma_g$ (or, equivalently, $\tau_g$)
is governed by a second-order differential equation of the type
$d^2 \sigma_g / dt^2 \simeq - \alpha_g^2 \sigma_g /4 $, supplemented
with a drift term ($R_g$) and a noise term ($\eta_g$) in both the
field $\sigma_g$ and its velocity $d\sigma_g / dt$.
\par

The detail of the Monte Carlo algorithm for implementing the
complex Langevin simulation will be reported elsewhere.
In short, the Langevin algorithm proceeds essentially
in two alternating steps: (i) for the current value of the fields,
calculate $O$ and store it; (ii) update the fields by calculating all their
time-derivatives $d\sigma_g /dt$, using expression (\ref{eq_R}) for 
estimating the $R_g$'s. The time-average of $O$
then yields the output of the quantum computation, the statistics being
controlled by adjusting the length of the random walk.
\par

\section{Example: the quantum FFT circuit}
Let us now discuss the scaling of the computational effort
required to simulate a quantum algorithm, focusing on
the quantum FFT algorithm~\cite{bib_coppersmith} used in Shor's
factoring algorithm as an illustration.\footnote{Note that the FFT
is not the most computationally demanding task in Shor's algorithm,
but this is unimportant for our illustrative purpose here.}
Since Shor's algorithm
has been described in details in the literature (see, {\it e.g.},
\cite{bib_ekertjozsa} for a review), it will be sufficient to note
that, after a certain number of computational steps, the quantum
register is in a {\it periodic} superposition of states $|a\rangle$
labeled by an integer between 0 and $2^L-1$, the period being related to
the sought factor of the composite number. The register is then subjected
to a quantum FFT, resulting in a probabilistic estimate of the period
(the probability of success can be made arbitrarily close to one
by repeating the computation). The time-demanding task in the Monte Carlo
simulation of the quantum FFT is the update of the $4G$ auxiliary fields.
Performing one step of the random walk in auxiliary-field space
needs the computation of $4G$ time-derivatives,
requiring each the calculation of a single [{\it cf.} 
Eqs.~(\ref{eq_dUdsigma1}) and (\ref{eq_dUdsigma2})]
one-bit matrix element (involving a product of about $G/L$ non-unit
$2\times2$ matrices).
Thus, since $G$ scales as $L^2/2$ for the quantum FFT circuit,
of the order of $L^3$ computation steps ($2\times2$ matrix multiplication)
are necessary to perform one step of the random walk.
Assuming that the number of steps necessary to achieve
a given statistical error in the estimate of $\langle O \rangle$
does not grow exponentially with $G$ (the sign problem should be overcome
and the auto-correlation time of the random walk
should not be exponential in $G$), 
the total number of computation steps would be polynomial in $L$.
This does not rule out the possibility that,
for a general quantum algorithm, the simulation effort might be
{\it polynomial} in $L$ whenever the number of gates $G$
required in the quantum circuit is polynomial in $L$. This is
an open question.
\par

\begin{figure}
\caption{Two-bit quantum Fast Fourier Transform circuit. It requires
two 1-bit Hadamard gates and one 2-bit controlled-phase gate.}
\vskip 0.25cm
\centerline{\psfig{figure=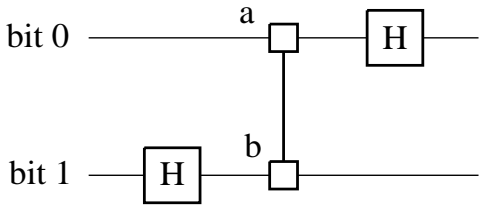,width=2.5in,angle=-90}}
\label{fig_fft}
\vskip -0.25cm
\end{figure}

As an example, we consider here a two-bit quantum FFT, {\it i.e.}, the
quantum computation of the discrete Fourier transform of a 
4-point function (see Fig.~\ref{fig_fft}).
The input qubits of the quantum register
($L=2$) are labeled 0 (and 1) for the least (and most) significant qubit.
The two-bit quantum FFT circuit~\cite{bib_coppersmith} requires
a single two-bit gate, a controlled-phase operator $C_{01}=e^{i\omega AB}$
acting on qubits $a$ and $b$,
with $\omega=\pi/2$ and $A=B=(1-\sigma_z)/2$,
and two additional one-bit gates $H_0$ and $H_1$, with H being the
Hadamard transformation,
\begin{eqnarray}
H|0\rangle &\to& (|0\rangle+|1\rangle)/\sqrt{2} \;,  \nonumber \\
H|1\rangle &\to& (|0\rangle-|1\rangle)/\sqrt{2} \;.
\end{eqnarray}
The total unitary transformation is the ordered product $U=H_0 C_{01} H_1$.
The two one-bit gates $H_0$ and $H_1$ can be
incorporated into the two-bit gate, which can in turn be 
written in terms of field-dependent one-bit gates
using the Hubbard-Stratonovich
representation, yielding $U[\sigma] = U_0(\sigma) U_1(\tau)$, with
\begin{eqnarray}
U_0(\sigma)&=&{1\over \sqrt{2}} \left(  \begin{array}{cc}
1 & e^{i\omega\sigma} \nonumber \\
1 & -e^{i\omega\sigma}
\end{array} \right) \\ 
U_1(\tau)&=&{1\over \sqrt{2}} \left(  \begin{array}{cc}
1 & 1 \\
e^{i\omega\tau} & -e^{i\omega\tau}
\end{array} \right)
\end{eqnarray}
For a simple test of the Langevin algorithm, we consider here
the Fourier transform of a constant function, {\it i.e.} the initial state
is the product state 
$2^{-1/2} (|0\rangle + |1\rangle) \otimes 2^{-1/2} (|0\rangle + |1\rangle)$.
The complex action can then be simply expressed as
\begin{equation}
S = \omega (\sigma\tau - \sigma' \tau') 
+ i \ln \left( { 1+e^{i \omega (\sigma-\sigma')}  \over 2} \right) \;,
\end{equation}
depending on the 4 auxiliary fields $\sigma$, $\tau$, $\sigma'$, and
$\tau'$. A straightforward calculation shows that the stochastic differential
equations obeyed by the fields are
\begin{eqnarray}
\frac{d\sigma}{dt} &=& - i {\omega \over 2} (\tau-1/2) 
- {\omega \over 4} \tan\left({\omega\over 2} (\sigma-\sigma') \right) 
+\eta_{\sigma}\;, \nonumber \\
\frac{d\sigma'}{dt} &=& i {\omega\over 2} (\tau'-1/2) 
+ {\omega\over 4} \tan\left({\omega\over 2} (\sigma-\sigma') \right) 
+\eta_{\sigma'}\;,  \nonumber \\
\frac{d\tau}{dt} &=& - i {\omega\over 2} \sigma  +\eta_{\tau} \;, \nonumber \\
\frac{d\tau'}{dt} &=& i {\omega\over 2} \sigma'  +\eta_{\tau'}
\end{eqnarray}
The Monte Carlo simulation of these equations is easy to perform. 
(Note that the fixed point of the Langevin dynamics---{\it i.e.},
the path of minimum action---is $\sigma=\sigma'=0$, $\tau=\tau'=1/2$.)
The resulting Monte Carlo averages for the 1-bit observables
$O_0 = |0\rangle \langle 0|$ and  $O_1 = |0\rangle \langle 0|$
converge to 1, implying that the expectation value for the output register
is $|00\rangle$, as expected (the spectrum has a continuous component only).
The simulation of larger quantum circuits using this technique is the
subject of further work to be reported elsewhere.

\section{Conclusion}
We have shown that a quantum computer can be treated as a genuine
quantum many-particle system, and that this approach sheds new light
on quantum computation.
More specifically, the use of a quantum Monte Carlo method might 
be interesting when considering ``large'' quantum computers
because of the polynomial scaling of the auxiliary-field space
in the dimension of the problem. This advantage, however, is conditional
on a circumvention of the Monte Carlo ``sign'' problem. In this respect,
the use of a Langevin algorithm as a possibly efficient simulation technique
is discussed. The stochastic simulation of quantum computation proposed
here could be useful for at least two reasons:
(i) it could help in devising actual quantum
computers by avoiding the need for an explicit experimental
realization to test a quantum algorithm; (ii) it could give rise
to a new class of ``quantum-inspired'' algorithms that could be
implemented on an ordinary classical computer for solving
certain computationally hard problems.
\par
\bigskip
 
We acknowledge C. Adami for many helpful discussions.
This work has been funded in part by the NSF under Grant
Nos. PHY 94-12818 and PHY 94-20470, and by
a grant from DARPA/ARO through the QUIC Program (\#DAAH04-96-1-3086).

\end{document}